\documentclass[structabstract]{aa}
\usepackage{txfonts}
\usepackage{graphicx}
\usepackage{amsbsy}

\usepackage{natbib}
\bibpunct{(}{)}{;}{a}{}{,}

\begin{document}

\title{Mass distribution in an assembling super galaxy group at $z=0.37$}
\author{Merijn Smit \inst{1}\thanks{\email{msmit@strw.leidenuniv.nl}} \and Tim Schrabback \inst{2,1,3} \and Malin Velander \inst{1} \and Konrad Kuijken \inst{1} \and Anthony H. Gonzalez \inst{4} \and John Moustakas \inst{5} \and Kim-Vy H. Tran \inst{6}}
\institute{Leiden Observatory, Leiden University, PO Box 9513, 2300RA Leiden, The Netherlands \and Argelander Institute for Astronomy, University of Bonn, Auf dem H\"ugel 71, D-53121 Bonn, Germany \and Kavli Institute for Particle Astrophysics and Cosmology, Stanford University, 382 Via Pueblo Mall, Stanford, CA 94305-4060, USA \and Department of Astronomy, University of Florida, Gainesville, FL 32611, USA \and Department of Physics and Astronomy, Siena College, 515 Loudon Road, Loudonville, NY 12211, USA \and George P. and Cynthia W. Mitchell Institute for Fundamental Physics and Astronomy, Department of Physics \& Astronomy, Texas A\&M University, College Station, TX 77843, USA}
\date{Received <date> / Accepted <date>}

\abstract
{}
{We present a weak gravitational lensing analysis of supergroup SG1120$-$1202, consisting of four distinct  X-ray-luminous groups that will merge to form a cluster comparable in mass to Coma at $z=0$. These groups lie within a projected separation of 1 to 4 Mpc and within $\Delta v=550$ km s$^{-1}$ and form a unique protocluster to study the matter distribution in a coalescing system.}
{Using high-resolution {\em HST}/ACS imaging, combined with an extensive spectroscopic and imaging data set, we studied the weak gravitational distortion of background galaxy images by the matter distribution in the supergroup. We compared the reconstructed projected density field with the distribution of galaxies and hot X-ray emitting gas in the system and derived halo parameters for the individual density peaks.}
{We show that the projected mass distribution closely follows the locations of the X-ray peaks and associated brightest group galaxies. One of the groups that lies at slightly lower redshift ($z\approx 0.35$) than the other three groups ($z\approx 0.37$) is X-ray luminous, but is barely detected in the gravitational lensing signal. The other three groups show a significant detection (up to $5 \sigma$ in mass), with velocity dispersions between $355^{+55}_{-70}$ and $530^{+45}_{-55}$ km s$^{-1}$ and masses between $0.8^{+0.4}_{-0.3} \times 10^{14}$ and $1.6^{+0.5}_{-0.4}\times 10^{14} h^{-1} M_{\odot}$, consistent with independent measurements. These groups are associated with peaks in the galaxy and gas density in a relatively straightforward manner. Since the groups show no visible signs of interaction, this supports the hypothesis that we observe the groups before they merge into a cluster.}
{}


\maketitle

\nocite{Gonzales05}
\nocite{Tran08}
\nocite{Tran09}
\nocite{KK06}
\nocite{KSBp_Erben}

\section{Introduction}

In the framework of hierarchical structure formation \citep{Peebles70}, matter overdensities grow through merging and accretion from the scales of galaxies up to those of large-scale structure (LSS). In the concordance $\Lambda$CDM cosmology, the large-scale structure of the Universe is driven by the density fluctuations of dark matter, which provide the initial framework for subsequent structure formation. As such, the mass distribution in the Universe is the driving force behind the formation of clustering sites for astrophysical processes, such as galaxy groups and clusters.

In galaxy formation and evolution, environment plays a role of major importance. Most galaxies are found in groups and clusters \citep[e.g.,][]{2PIGG}, and observations indicate that the main part of galaxy evolution takes place in the group environment, with significant post-processing occurring in clusters \citep[][hereafter T08 and T09]{Tran08,Tran09}. A detailed understanding of the total mass (dark and visible) and the structure of the mass density distribution is therefore necessary to understand both the processes of group and cluster formation and fundamental scaling relations \citep[e.g., ][]{COSMOS,Hoekstra_CCCP,Linden2014} as well as to distinguish the latter from intrinsic variances in astrophysical processes.

Most overdensities are detected using visible, that is, baryonic means. Common methods use galaxies \citep[red sequence and spectroscopic association, e.g.,][]{2PIGG,RCS05} or gas \citep[X-ray emission or the SZ effect, e.g.,][]{SS70,SS72,Finoguenov07}. While these methods are efficient, they might not always be as effective: they rely on the presence of baryonic matter, while the matter distribution is driven by dark matter. Furthermore, the subsequent classification relies on observing the results of complex (astrophysical) processes, which introduces a significant intrinsic scatter in properties such as X-ray temperatures, star formation rate, and galaxy morphologies in different structures of comparable mass.

Gravitational lensing is the only direct probe of the total mass distribution, in the sense that it does not rely on astrophysical assumptions. A lower signal-to-noise ratio (S/N) makes it a less efficient detection method except for massive structures, but in combination with complementary methods, it is a powerful independent tool. From a statistical perspective, as an independent, direct measurement, it can serve as a calibrator for mass-observable scaling relations \citep[e.g., ][]{COSMOS,Hoekstra_CCCP,Linden2014}. In individual systems it provides an independent estimation of the (projected) density field and can shed light on aspects such as interaction, dynamical state, and offsets between the overall matter distribution and the baryonic matter.

Direct reconstructions of the density fields of individual systems do not have a high resolution and are predicted to show significant noise fluctuations \citep{Waerbeke2000}. However, it gives an important qualitative indication of the dominant density distribution, independent of the presence of baryons. To determine the parameters of the matter distribution, either in a statistical survey or for individual systems, robust centroiding is an important task. Using other tracers of the center of mass, such as X-ray detections, brightest cluster/group galaxies (BCG/BGG), or the luminosity-weighted mean (LWM) position, gravitational lensing can significantly constrain halo masses and concentrations \citep[e.g.,][]{George12}.

Galaxy clusters and, in the past decade, galaxy groups, are now identified in a robust manner, including examples of accretion of smaller structures onto existing clusters. However, we have less observational evidence of the connection between structures on various scales, that is to say, the initial assembly of clusters from groups and galaxies. In this study, we perform a weak lensing analysis of SG1120$-$1202 \citep[][hereafter G05]{Gonzales05}, an assembling system of four galaxy groups at $z\sim 0.37$ discovered in the Las Campanas Distant Cluster Survey \citep[LCDCS, ][]{Gonzales01}. These groups are gravitationally bound and will merge into a galaxy cluster comparable in mass to Coma by $z=0$. The supergroup, hereafter SG1120, is confirmed by X-ray imaging and optical spectroscopy and has already formed a red sequence \citep[see, e.g., G05, T08, T09, and][]{Kautsch08,Just11,Freeland11}.

The individual subgroups are in the low-mass regime of X-ray groups, $M_{200} \sim 10^{13}$ to $10^{14}M_{\odot}$ and $\sigma_\mathrm{v} \sim 400$ km s$^{-1}$, and have not yet interacted. The aim of this study is to determine the total matter distribution in the system (dark and baryonic) and to constrain individual halo masses.

This paper is organized as follows. We summarize the general framework for weak lensing in Sect. 2. In Sect. 3 we briefly describe the data we use, while Sect. 4 covers the framework of measurement and analysis methods. In Sect. 5 we discuss the results and the scientific implications. Section 6 gives a summary of our conclusions.

Throughout this paper we assume a Planck \citep{Planck13XVI} cosmology with $\Omega_{\mathrm{M}}=0.3183$, $\Omega_{\mathrm{{\Lambda}}}=0.6817$ and $H_0=67.04$ km s$^{-1}$ Mpc$^{-1}$.

\section{Weak lensing framework}

Gravitational lensing is the effect of curved space-time on the paths of light rays from distant sources to the observer as they pass through the potential of foreground structures. This geometrical effect leads to a displacement of point sources on the projected plane of the sky. The differential effect on extended sources leads to magnification and distortion effects. This is commonly described as a coordinate transformation \begin{equation} \left( \begin{array}{c} x' \\ y' \end{array} \right) =\left( \begin{array}{cc} 1-\kappa-\gamma_1 & -\gamma_2 \\ -\gamma_2 & 1-\kappa+\gamma_1 \end{array} \right) \left( \begin{array}{c} x \\ y \end{array} \right) \,, \end{equation} where the trace component $\kappa$ is known as the convergence and the reduced symmetric part is determined by the gravitational shear $(\gamma_1,\gamma_2)$.

Since we do not know the intrinsic source sizes or magnitudes, we can only measure the net {\em \textup{distortion}} or {\em \textup{reduced shear}} $(g_1,g_2) \equiv (\gamma_1,\gamma_2) / (1 - \kappa)$: \begin{equation} \left( \begin{array}{c} x' \\ y' \end{array} \right) = \left( 1-\kappa \right) \left( \begin{array}{cc} 1-g_1 & -g_2 \\ -g_2 & 1+g_1 \end{array} \right) \left( \begin{array}{c} x \\ y \end{array} \right) \,, \end{equation} where the transformation is written as a multiplication of $( 1-\kappa )$, which we do
not identify, and a distortion matrix describing the alignment of lensed sources in the foreground potential.

The observed shape of a background source is not a pure tracer of the gravitational lensing effect, but the combined effect of an intrinsic galaxy shape and any distortion of that shape, including gravitational lensing. Systematic effects such as telescope aberrations and detector systematics likewise contribute, which need to be corrected for.  

The uncertainties the intrinsic shapes introduce is referred to as shape noise, and the amount of background sources available determines the precision of the results. The shape noise averages out statistically if we assume that the background sources are randomly oriented intrinsically. If faint members of the foreground structure are not identified and removed from the sample of background sources, these shapes will not be affected by gravitational lensing, but might be aligned with the potential of the structure under investigation. These effects are known as intrinsic alignment \citep[see, e.g.,][]{Mintr}. However, recent results suggest that intrinsic alignments should have negligible influence for current cluster weak lensing studies \citep{Sifon2014}.

The average measured distortion, corrected for systematic effects, can then be related to the projected density distribution in the lensing structure through \begin{equation} \kappa(\boldsymbol{\theta}) = \frac{4\pi G}{c^2} \Sigma(\boldsymbol{\theta}) \frac{D_{\mathrm{ol}}D_{\mathrm{ls}}}{D_{\mathrm{os}}} \,, \end{equation} where ${\boldsymbol{\theta}}$ represents the angular coordinates on the plane of the sky, $\Sigma(\boldsymbol{\theta})$ is the projected density distribution, and $D_{\mathrm{i}}$ are the angular diameter distances between the observer, lens, and background sources (luminosity distances, sometimes written as $D_{\mathrm{l}}$, are not used throughout this paper).

Normalized by $4\pi G c^{-2}$, the convergence $\kappa$ is therefore also known as the dimensionless surface mass density, directly related to the lensing density distribution and the lensing geometry. For axisymmetric lenses, $|\gamma|(\theta) = \bar\kappa(<\theta) - \kappa(\theta)$ with $\bar\kappa(<\theta)$ the mean surface mass density within a radial separation $\theta = |\boldsymbol{\theta}|$ to the lens centroid.

\section{Data}

For this project we made use of results of an extensive multiwavelength data set (see, e.g., G05, T08, and T09).

Key to our lensing analysis are optical data, consisting of high-resolution {\em HST}/ACS\footnote{Based on observations made with the NASA/ESA Hubble Space Telescope, obtained at the Space Telescope Science Institute, which is operated by the Association of Universities for Research in Astronomy, Inc.} imaging used for shape measurements, as well as {\em VLT}/VIMOS \citep{LeFevre2003}, {\em VLT}/FORS2 \citep{Appenzeller1998}, and {\em Magellan}/LDSS3 spectroscopy.

We also used the X-ray temperatures based on {\em Chandra}/ACIS imaging and stellar masses inferred from {\em VLT}/VIMOS $BVR$ photometry (T08) and complement our optical color information with {\em KPNO}/FLAMINGOS near-infrared (NIR) $K_s$ imaging. 

We use $\alpha=11\fh 19\fm 58\fs 0$, $\delta=12\degr 03\arcmin 33\farcs 0$ as center of coordinates, which roughly is the center of the VIMOS imaging data. To convert angular to physical separations, we use a reference redshift of $z=0.37$; this is the median of the redshifts of the four BGGs.

\begin{figure}[h]
\centering
\resizebox{\hsize}{!}{
\includegraphics[angle=90]{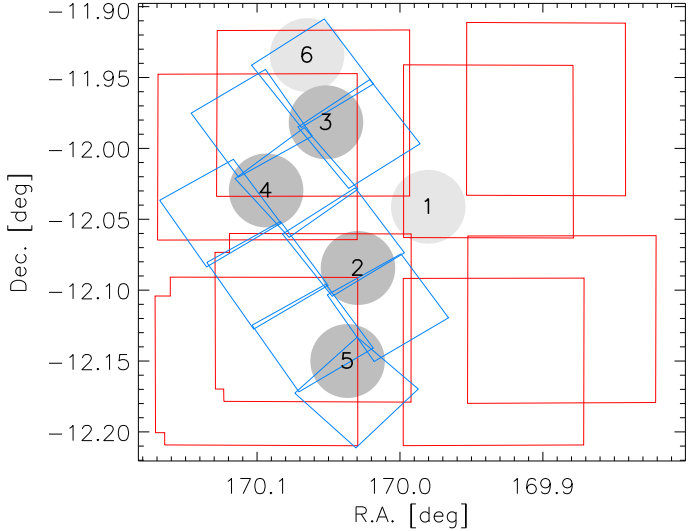}
}
\caption{Layout of the {\em VLT}/VIMOS pointings (red) and {\em HST}/ACS pointings (blue). The detected X-ray peaks are shown as well (gray), with the radius of the circles 0.5 $h^{-1}$ Mpc. The X-ray peaks 1 and 6 (light gray) are associated with structures at higher redshift, beyond SG1120 (G05).}
\label{layout}
\end{figure}

\subsection{HST imaging}

The {\em HST}/ACS imaging data were taken in July 2005 and January 2006 and consist of ten pointings, forming a contiguous $11'\times 18'$ mosaic. Each tile was observed in F814W ($0\farcs 05$/pixel) for 2ks over four dithered exposures. Figure \ref{layout} shows the layout of the different pointings.

We reduced the data with the same pipeline as employed in \citet{TS10}, which uses \texttt{MultiDrizzle} \citep{Koekemoer03} to stack exposures and remove cosmic rays. It also includes careful refinement of shifts and rotations between exposures as well as optimized weighting.

\citet{TS10} found that using \texttt{MultiDrizzle} with the default cosmic-ray rejection parameters can cause central stellar pixels to be flagged as cosmic rays, especially when there are significant PSF variations between exposures. Galaxies are not affected, due to their shallower light profiles. To avoid differences in the effective stacked PSFs, we created separate stacks for stars and galaxies, with less aggressive cosmic-ray rejection for the former.

For a more detailed description of the reduction process, we refer to \citet{TS10}.

\subsection{Spectroscopy}

\begin{figure}[h]
\centering
\resizebox{\hsize}{!}{
\includegraphics[angle=90]{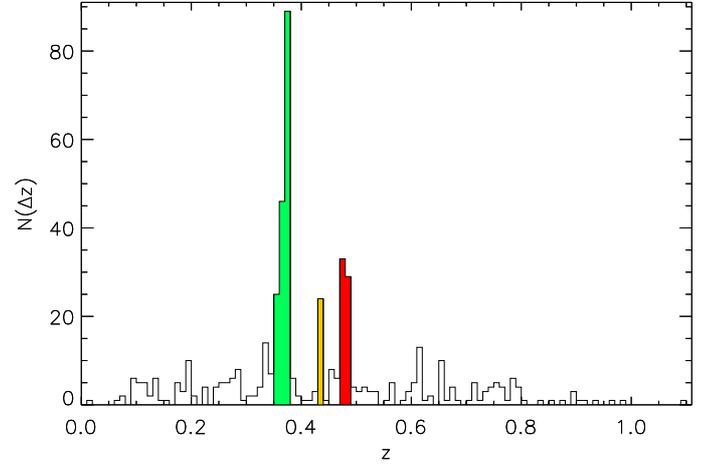}
}
\caption{Redshift distribution of the spectroscopic targets around SG1120. Three significant peaks are identified between $0.35<z<0.37$, associated with SG1120; between $0.43<z<0.44$, unassociated but concentrated slightly north of peak 1; and between $0.47<z<0.49$, associated with X-ray peaks 1 and 6.}
\label{specdist}
\end{figure}

We employed optical spectroscopy consisting of three subsets of data. The first subset of targets was selected from a magnitude-limited catalog ($R\le 22.5$), with preference given to objects in visually overdense regions (G05), and observed using VIMOS. Follow-up spectroscopy was selected from $K_s$ catalogs ($K_s\le 20$) and carried out on LDSS3 and FORS2. Figure \ref{specdist} shows the redshift distribution of the final target selection, using a redshift quality cut as defined in T08.

\begin{figure}[h]
\centering
\resizebox{\hsize}{!}{
\includegraphics[angle=90]{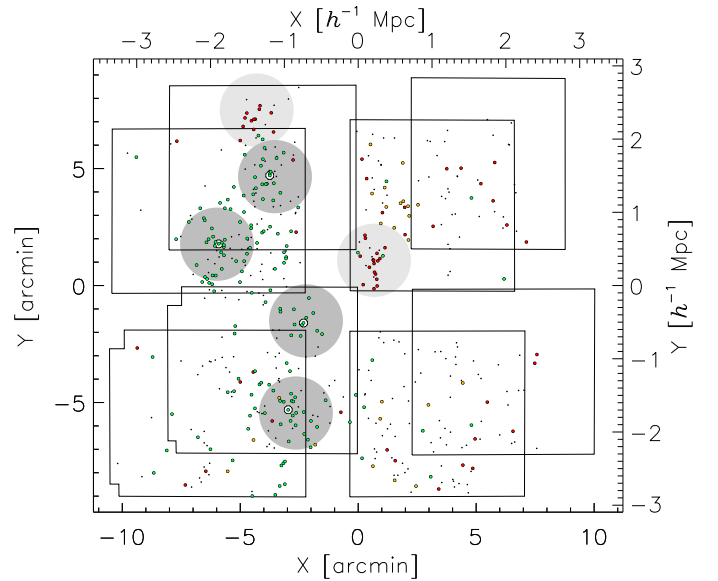}
}
\caption{Layout of spectroscopic targets, overlaid with the {\em VLT}/VIMOS pointings. $(x,y)=(0,0)$ corresponds to $\alpha=11\fh 19\fm 58\fs 0$, $\delta=12\degr 03\arcmin 33\farcs 0$. Colors correspond to the peaks in Fig. \ref{specdist}. The BGGs are indicated by larger circles.}
\label{specmap}
\end{figure}

Members of the subgroups of SG1120 were initially selected as galaxies within 500 kpc of their respective X-ray peaks within the redshift range $0.32 \le z \le 0.39$. We narrowed the redshift range to $0.34 \le z \le 0.38$ without the loss of any members. Figure \ref{specmap} shows a layout of the targets.

\subsection{Subgroups}

\begin{table}[h]
\caption{Properties of galaxy groups in SG1120 and the two structures identified at higher redshift.}
\label{tab_prop}
\centering
\begin{tabular}{cccccc}
\hline
\hline
BGG   & X-ray & z & $T$   & $\sigma_z$   & $N$ \\
ID    & peak  &   & (keV) & (km s$^{-1}$) &     \\
\hline
1      & 2 & $0.3522$ & $2.2^{+0.7}_{-0.4}$ & $303\pm 60$  & 13 \\ 
2      & 3 & $0.3707$ & $1.7^{+0.5}_{-0.3}$ & $406\pm 83$  & 19 \\ 
3      & 4 & $0.3712$ & $1.8^{+1.2}_{-0.5}$ & $580\pm 100$ & 29 \\ 
4      & 5 & $0.3694$ & $3.0^{+1.2}_{-1.0}$ & $567\pm 119$ & 21 \\ 
\hline                                                           
       & 1 & $0.4794$ & $2.3^{+0.4}_{-0.3}$ & $820\pm 101$ & 19 \\ 
       & 6 & $0.4801$ & \dots            & \dots & 12 \\ 
\hline                                                           
\end{tabular}                                                    
\end{table}

In Table \ref{tab_prop}, we give an overview of the properties of each subgroup, also given in G05, T08 and T09. We use the same numbering convention for the X-ray peaks and BGGs as these papers. Subtle differences in the number of group members are due to using the same selection criteria as T09, but a slightly different cosmology.

\begin{figure}[h]
\centering
\resizebox{\hsize}{!}{
\includegraphics[angle=90]{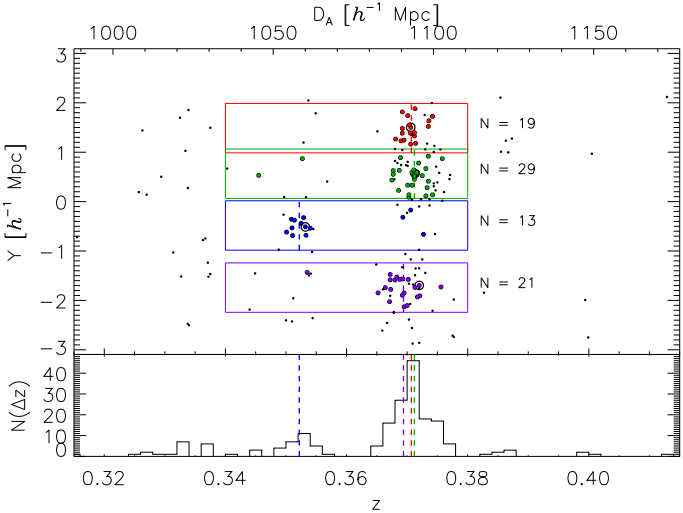}
}
\caption{Radial $(z,y)$ distribution of all objects (black dots, with a redshift quality of 3 as defined in T08). Objects within 500 kpc of an X-ray peak and the corresponding selection criteria are shown in different colors for easy distinction. The BGGs are indicated by larger circles, group redshifts by dashed lines.}
\label{zcuts}
\end{figure}

Figure \ref{zcuts} shows the radial $(z,y)$ distribution of galaxies within 500 kpc of each subgroup, with $0.315 \le z \le 0.415$. The subgroup associated with X-ray peak 2 and BGG 1 is at a slightly lower redshift, with an estimated $30\sim 40$ Mpc in angular diameter distance to the median redshift of the supergroup.

\section{Analysis}

In this section we briefly describe our method of shape measurement. We discuss the redshift distribution and selection of background sources. After establishing a reliable background catalog with robust shapes, we describe how we obtain a qualitative reconstruction of the projected mass density and complement this with density profiles to the subgroups based on the BGG and X-ray peak positions.

\subsection{KSB+ shape measurements}

The art of measuring accurate galaxy shapes is an ongoing field of investigation, as witnessed, for instance, by the Shear TEsting Programmes and the GRavitational lEnsing Accuracy Testing \citep[][hereafter STEP, STEP2, and GREAT08, GREAT10, and GREAT3]{STEP,STEP2,GREAT08,GREAT10,GREAT3}. We make use of the KSB method \citep{KSB}, the most commonly used and tested technique in the past decade, and discuss its application to ACS data.

For this study we used the same approach as \citet[][the TS pipeline in STEP and STEP2]{TS07,TS10} based on the implementation by \citet{KSBp_Erben}. KSB uses the first-order effects of distortions induced by gravitational shear and PSF on the weighted second moments of the light distribution of a source to estimate the reduced shear. We detect objects in the same way using {\tt SExtractor} \citep{SEX}.

We defined individual weights for this method based on the variance of the shear estimators from this pipeline as \begin{equation} w\propto\left(\sigma_{sn}^2 + \sigma_{e_1}^2 + \sigma_{e_2}^2\right)^{-1} \,, \end{equation} where we assumed a minimum variance of $\sigma_{sn}^2$, the intrinsic shape noise. Based on the findings in the STEP analyses, we expect an underestimation of the shears by KSB+ of about a few percent. We applied the same empirical correction factor as \citet{TS10} to account for this expected bias. 

The systematic distortion effects due to telescope and optic system give rise to shapes convoluted by a point spread function (PSF). The main source of variations of the ACS PSF is caused by changes in the telescope focus, causing spatial and temporal fluctuations \citep[see, e.g.,][]{TS07,Rhodes2007}.

A common strategy is to map the distortions caused by the PSF using the shapes of foreground stars, but the average number of stars in our ACS images is $\sim 20-40$. This leads to a poorly sampled PSF and an imperfect correction, causing significant residual distortions, especially detectable at the edges of the images where the tiles overlap slightly. We therefore adopted the same strategy as \citet{TS10} based on a principal component analysis of the PSF variation in dense stellar fields.

Furthermore, deterioration of the ACS CCDs over time due to constant exposure to cosmic rays in space leads to an effect called charge-transfer
inefficiency (CTI), causing charge trails in the CCD readout direction \citep[e.g.,][]{Rhodes2007,Massey2010,TS10}. These effects will affect the measured shear pattern and the reconstruction of the projected density distribution, and it is therefore important to correct for them.

Here we applied the same parametric CTI model as described in \citet{TS10} for the correction of the KSB+ polarizations.

After constructing the shape measurement catalogs, we applied several common selection criteria and cuts. These criteria are based on simulations and quality flags of the  detection and shape measurement pipelines, and they depend on the noise properties, on the variance and convergence of the model fits, and on the object and PSF size.

A list of the various selection criteria can be found in the appendix. Sources that pass the criteria of all pipelines number 7012, for a source density of $\sim 64$ galaxies/arcmin$^2$ with {\tt MAG\_AUTO} magnitudes $i_{814}<27.1$.

\subsection{Redshift distribution}

We acquired spectroscopic redshifts for 497 objects in our catalogs. The spectroscopic targets were selected based on magnitude, and preference was given to visually overdense regions, which means
that these spectroscopically confirmed members do not give a complete picture of the galaxy distribution in SG1120. The brightest confirmed supergroup member has $i_{814}=17.5$, while the spectroscopic survey remains $> 50\% $ complete to $i_{814}=20.5$ (T09). We find that confirmed supergroup members have numbers peaking between magnitudes $19.5 < i_{814} < 20.0$.

\begin{figure*}[h]
\centering
\resizebox{\hsize}{!}{
\includegraphics[angle=90]{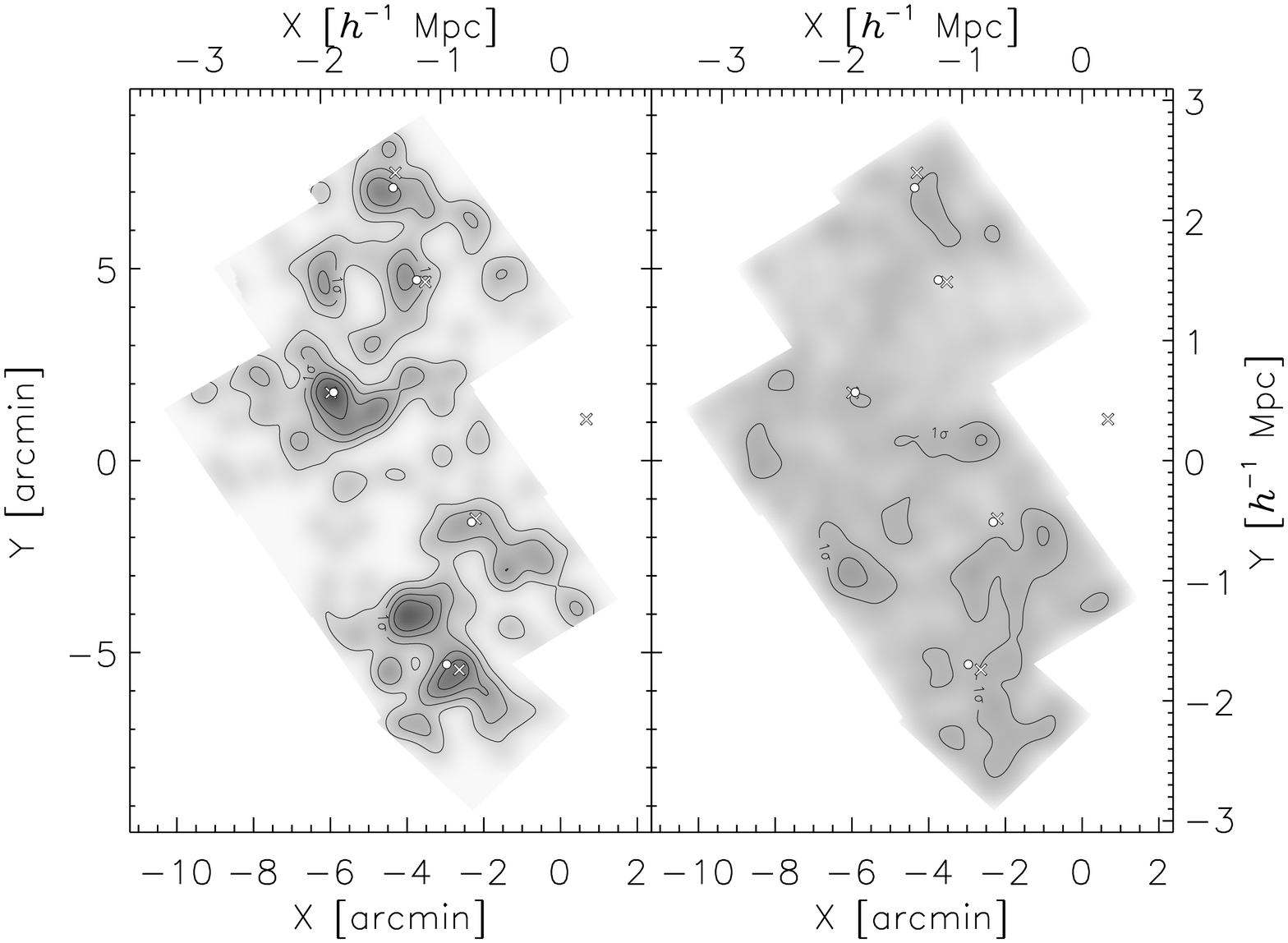}
}
\caption{Number density of sources with $i_{814} < 22$ {\em (left)} and $i_{814} > 22$ {\em (right)}, smoothed using a Gaussian smoothing kernel with a width of $20"$. Dense regions are shown as dark in a normalized grayscale. Contours correspond to fluctuations in integer standard deviations in number density.}
\label{ndens}
\end{figure*}

To separate background and foreground sources, we considered that group members are expected to dominate number counts in the magnitude range of confirmed members, while background number counts dominate at fainter magnitudes. We initially selected background sources as objects with $i_{814} \ge 22$ and assessed possible contamination by faint foreground objects. Figure \ref{ndens} shows the number density of sources with $i_{814} < 22$ and $i_{814} > 22$, where we have used a Gaussian smoothing with a $20"$ kernel width.

\begin{figure}[h]
\centering
\resizebox{\hsize}{!}{
\includegraphics[angle=90]{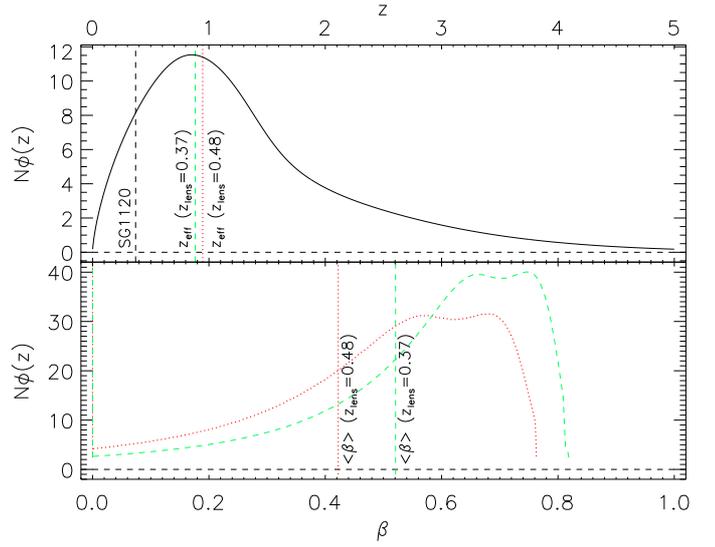}
}
\caption{Parametric redshift distribution of sources with $i_{814} > 22$ {\em (upper panel)} and the corresponding distribution in lensing efficiency $\beta$ {\em (lower panel)}. In the upper panel, the dashed green line corresponds to $z_{\mathrm{eff}} \approx 0.88$ with respect to SG1120 and the dotted red line corresponds to $z_{\mathrm{eff}} \approx 0.95$ with respect to the two structures at higher redshift. In the lower panel, the dashed green curve shows the distribution in $\beta$ with respect to SG1120, with $\left\langle\beta\right\rangle \approx 0.52$ (dashed green vertical line), and the dotted red curve the distribution in $\beta$ for the two structures at $z=0.48$, with $\left\langle\beta\right\rangle \approx 0.42$ (dotted red vertical line).}
\label{lenseff}
\end{figure}

Because gravitational lensing is a geometric effect that has a non-linear dependence on redshift, we took the expected redshift distributions into account, following the same parametrization as in \citet{TS10}. We show the total redshift distribution of sources with $i_{814} > 22$ in Fig. \ref{lenseff}. For a given lens redshift, such as in this particular system, the lensing signal has a linear dependence on the {\em \textup{lensing efficiency}} $\beta = \mathrm{max} \left\{ 0,D_{\mathrm{ls}}/D_{\mathrm{os}} \right\} $. We can therefore determine a mean lensing efficiency $\left\langle\beta\right\rangle$ for the sources with respect to each subgroup redshift.

As mentioned earlier, both X-ray peak 1 and 6 (G05) are associated with structures at higher redshift (both $0.46 \la z \la 0.48$). We must take the gravitational distortions caused by these background structures into account when trying to isolate the signal from SG1120. We therefore also determined a mean lensing efficiency for these structures.
 
We found average lensing efficiencies of $\left\langle\beta\right\rangle \approx 0.52$ for SG1120, corresponding to an effective background redshift of $z_{\mathrm{eff}} \approx 0.88$, and $\left\langle\beta\right\rangle \approx 0.42$ for the two background structures, corresponding to an effective background redshift of $z_{\mathrm{eff}} \approx 0.95$.

\subsubsection{Foreground contamination}

An intrinsic redshift distribution of sources with $i_{814} > 22$ implies that some of these objects are faint foreground sources or members of the SG1120 structure. Based on our parametric redshift distribution, we estimate that $\sim 9\%$ of our background sources to lie in front of SG1120.

Foreground sources are not lensed by the groups. We account for this dilution effect by assigning  $\beta = 0$ to this part of the redshift distribution in our definition of the lensing efficiency above.

\begin{figure}[h]
\centering
\resizebox{\hsize}{!}{
\includegraphics[angle=90]{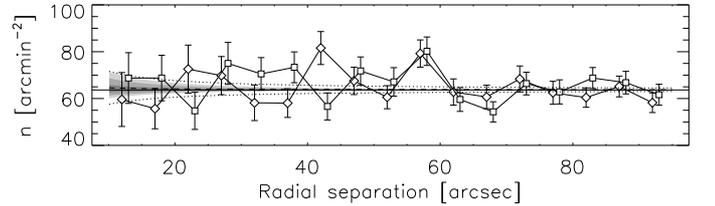}
}
\caption{Variations in galaxy number density $n$ as a function of radial distance from the lens positions, using the X-ray peaks (diamonds) and BGG positions (squares). Data points are slightly offset for clarity. Overplotted is the average number density of $\sim 64$ galaxies/arcmin$^2$ of the whole ACS mosaic (black solid line) and the best-fit radial profile (dashed) with $1\sigma $ errors (dotted). The estimated effect of the lensing magnification, $\mu^{\alpha -1} $, is shown in grayscale, varying the slope of the luminosity function between $0 < \alpha < 3$. Different shades in grayscale correspond to steps of $0.5\times 10^{14}$ in group mass $M_{200} $.}
\label{dens_prof}
\end{figure}

This assumes a random field of view, which is not the case for our observations, with known overdensities at $z\sim 0.37$ and $z\sim 0.48$. However, Fig. \ref{ndens} suggests no significant correlation between the distribution of these sources and the galaxy distribution of SG1120. To estimate possible variations in the number density $n$ of sources with $i_{814}>22$, we derived an average number density profile around the group centers, as shown in Fig. \ref{dens_prof}.

We used radial bins between $10"< \theta <95"$ to avoid the BGGs and the edges of the ACS coverage. We also considered only the group centers of groups 2 through 5, as group 6 is not entirely covered by the ACS pointings. Finally, we averaged the signal over all four subgroups to increase the signal-to-noise ratio.

We then quantified the radial dependence of the galaxy number density by fitting a parameterized profile given by $n(\theta ) = (1+a/\theta ) n_{\mathrm{bg}}$, with $\theta $ in arcseconds and with $n_{\mathrm{bg}} = 64/\mathrm{arcmin}^2$ fixed. (In fact, if we allow $n_{\mathrm{bg}}$ to vary, we recover $n_{\mathrm{bg}} = (64\pm 3 )/\mathrm{arcmin}^2$.) We found $a=0.14\pm 1.11$, consistent with no trend in galaxy number density with radial separation from the lens centers.

To interpret this radial number density profile, we have to consider both the presence of unidentified faint group members and the effect of the {\em \textup{lensing magnification}} $\mu $ \citep[see, e.g.,][and references therein]{Bartelmann01}. The presence of unidentified group members would increase the number density. Magnification increases both the observed flux of background sources, leading to an increase in $n$, and the solid angle behind the lenses, causing a dilution of $n$ (not to be confused with the dilution of the {\em \textup{shear signal}} caused by unlensed foreground objects in the background source sample). It then depends on the slope $\alpha $ of the luminosity function whether the lensing magnification causes a net increase or decrease in number density by $\mu^{\alpha -1}$, where $\mu $ and $\alpha $ depend of the source redshift. Both effects were shown by \citet{Hildebrandt09}. A decrease could cancel the effect of unidentified group members.

We wish to obtain a rough estimate of the expected influence of magnification to check whether it is smaller than the statistical uncertainty. For this we ignored the redshift dependencies of $\mu$ and $\alpha$ and considered a wide range $0<\alpha<3$, which was simply chosen to assess all possible variations in the magnification without making assumptions about the luminosity function. We used a group mass of $M_{200} = 1.0\times 10^{14}$, where $M_{200}$ is defined as the total mass within a radius of $r_{200}$ of a halo, where the mean density of the halo is $\bar\rho(<r_{200}) = 200\rho_{\mathrm{crit}}(z)$. For a robust estimate, we also considered variations in $M_{200}$ of $\pm 0.5\times 10^{14}$. The results are plotted in Fig. \ref{dens_prof} in grayscale.

We estimate possible magnification effects to be smaller than the statistical uncertainties from potential residual group member contamination. We expect that any residual {\it \textup{excess}} contamination by member galaxies of SG1120 in the source sample must be small and comparable to regular line-of-sight variations. In Sect. 5.2, we confirm that we do not need to apply a dilution correction for excess contamination from the supergroup itself.

\subsection{Lensing analysis}

Our approach to determining the matter distribution in SG1120 is twofold. First, we show that the distribution of light (galaxy number densities, BGGs, and X-ray peaks) is closely correlated with the underlying mass distribution. Second, we determine the density profile parameters for each subgroup, taking into account the effect of each subgroup and background structure simultaneously.

\subsubsection{Reconstruction of the mass distribution}

We used a Kaiser-Squires \citep[KS, ][]{KS93} inversion technique to reconstruct the surface mass density. We smoothed the data onto a rectangular grid, using a Gaussian smoothing kernel with a width of $20"$, equal to the smoothing used for the galaxy number densities in Fig. \ref{ndens}.

We investigated possible systematic errors in our data by changing the phase of the shear by $\frac{1}{2}\pi$, which corresponds to rotating the background galaxies by $\frac{1}{4}\pi$. The distortion caused by weak lensing does not introduce a curl in the shear field, and the resulting reconstructed map should display only noise in the absence of systematic errors. 

\subsubsection{Density profile parameters}

Earlier studies indicate that the groups are infalling for the first time and have not yet interacted, although X-ray measurements show a possible onset of interaction (G05). We considered the groups as individual overdensities with spherically symmetric density distributions and derived halo parameters for each group, including the background structure around X-ray peak 6.

We considered two types of density profiles and two possible choices of group centroids. We considered the Navarro-Frenk-White \citep[NFW, ][]{NFW} density profile and compared this to the singular isothermal sphere (SIS) model. 

The SIS profile is determined by a single free parameter, the halo velocity dispersion $\sigma_{\gamma}$, where the subscript $\gamma$ is used to distinguish this parameter, derived from a two-dimensional model of the projected mass density, from other derivations of velocity dispersion, such as the one-dimensional $\sigma_z$ derived from the redshift distribution. The advantages of this profile are its simplicity and the linear dependence of the lensing signal on the squared velocity dispersion. The tangential component of the shear with respect to the group center is given by \begin{equation} \gamma_\mathrm{t}(\theta) = \frac{2\pi}{c^2} \sigma_{\gamma}^2 \frac{\beta}{\theta} \,, \end{equation} where $\theta$ indicates the separation from the center in radians. This allows for a straightforward interpretation of any possible correlation between the fitted parameters of different subgroups.

The NFW profile is usually expressed in terms of its mass and concentration and depends on redshift. The halo mass $M_{200}$ is given by \begin{equation} M_{200} \equiv 200\rho_{\mathrm{crit}}(z) \frac{4}{3}\pi r_{200}^3 = 100\frac{H(z)}{G} r_{200}^3 . \end{equation} The concentration $c_{200}$ is defined as the relation between the characteristic shape of the density profile and $r_{200}$. The analytical formulas for the shear signal of an NFW profile can be found in \citet{Wright2000} and \citet{Bartelmann96}.

Because of the lower S/N, the centers of dark matter haloes should not be estimated directly from the lensing data when determining density profile parameters. Instead, one has to rely upon visible tracers such as peaks in the X-ray emission of hot gas or the brightest or heaviest galaxy (e.g., in terms of a stellar mass as derived in T08) in the group or cluster. If the fitted halo model is offset from the true underlying halo, the fit is inferior and the introduced systematic uncertainties can be significant \citep{George12}. In particular, the halo mass will on average be underestimated, while the uncertainties, most often determined from confidence levels, will be increased. This leads to both a biased and a less effective study.

As described in \citet{George12}, there are several choices possible as tracer of the halo center. These can be based upon a central galaxy, several or all of the associated galaxies, or the X-ray flux. In this study, the haloes under consideration are part of a coalescing system, and an offset from the true halo center of some or all of these tracers is not unlikely. However, the BGGs of the subgroup are also the most massive group galaxies \citep[MMGG,][]{George12} in terms of stellar mass and magnitudes in most observed bands and coincide well with the X-ray peaks (T08). We derived the parameter values using both options and determined whether these are consistent.

Given the close angular separation of the X-ray peaks, we did not compute azimuthally averaged profiles. Instead, we computed the total lensing distortion $\boldsymbol{g} = \sum \boldsymbol{g}_i$ for each background source induced by each of the six foreground structures. This is valid if we assume $g \ll 1$, which is certainly the case for the sources where the distortion is not dominated by one of the lensing structures.

We then determined profile parameters for each subgroup using a $\chi^2$ minimization. For X-ray peak 1, we assumed $\sigma = 820$ km s$^{-1}$ from G05 and an order of magnitude $M_{200} = 3.7\times 10^{14} h^{-1} M_{\odot}$ and assessed the effect of omitting the influence of this background structure.

\section{Results}                                                
                                                                 
In this section we discuss the reconstructed density distribution and best-fit profile parameters, and we show that SG1120 is consistent with expectations from hierarchical structure formation, even though the system is not relaxed.

\subsection{Matter distribution}                                 

\begin{figure*}[h]                                                
\centering                                                       
\resizebox{\hsize}{!}{                                           
\includegraphics[angle=90]{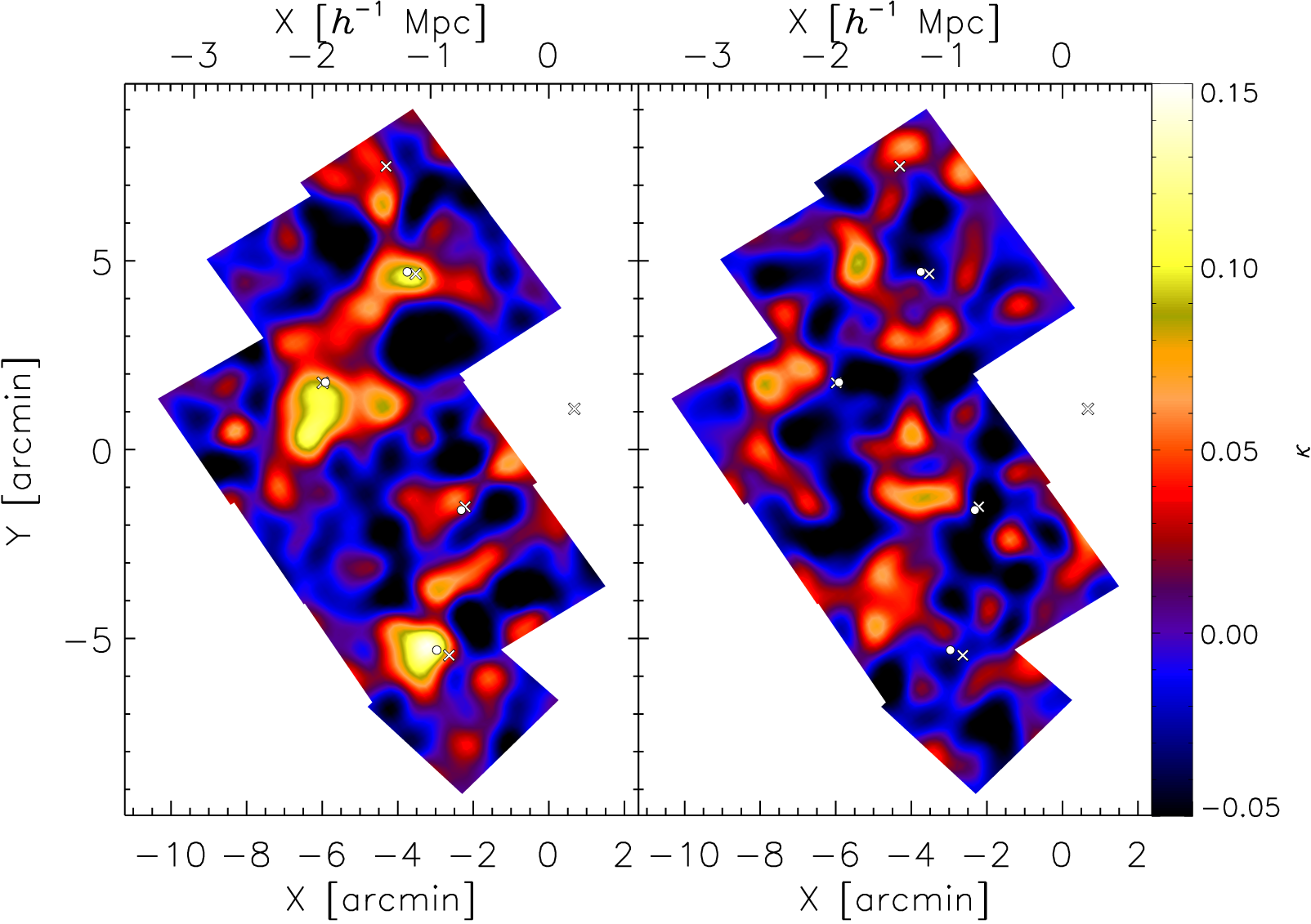}
}                                                                
\caption{Smoothed map of the reconstructed projected density distribution (left) and the imaginary control signal (right) where the shear signal is rotated out of phase. X-ray peak positions are indicated by white crosses and BGG positions by white circles.}
\label{ksinv}                                                    
\end{figure*}

In Fig. \ref{ksinv} we show the reconstruction of the projected surface mass density. We detect significant peaks near three of the foreground structures. We do not detect a significant peak in the density distribution near X-ray peak 2.

We considered the results of our mass reconstruction in a qualitative manner. The peaks in our surface mass density reconstruction coincide very well with the peaks in galaxy number density (Fig. \ref{ndens}) and X-ray emission (G05), within smoothing scales. We  found no significant `dark' overdensities either, and small offsets between peaks using various tracers are expected in a coalescing system. Finally, the map shows significantly stronger peaks than the control map.

\subsection{Individual groups}                                   

\subsubsection{SIS velocity dispersions}

\begin{figure}[h]
\centering
\resizebox{\hsize}{!}{
\includegraphics[angle=90]{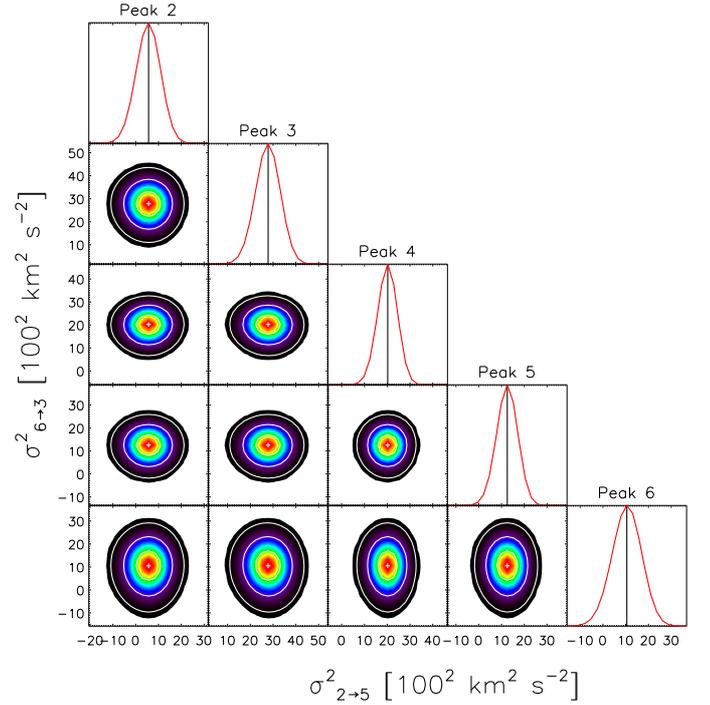}
}
\caption{Marginalized 2D $\chi^2$ distributions of the simultaneous fit to the individual subgroup velocity dispersions, together with the marginalized 1D likelihoods for each subgroup. Overplotted are the $68.3\%$,$95.4\%,$ and $99.7\%$ confidence levels.}
\label{Chi2}
\end{figure}

We present the results of the joint $\chi^2$ minimization fit of SIS profile parameters around the X-ray peaks in Fig. \ref{Chi2}. The reduced $\chi^2$ value is $\chi^2_{\nu}=1.4$.

The combined contours of Fig. \ref{Chi2} show no features that indicate significant degeneracies between the individual group $\sigma_{\gamma}$ values. While it is to be expected that nearby mass concentrations influence the shear pattern around an individual lens, we conclude that noise is a dominant factor in these results. More massive haloes or smaller halo separations can be expected to increase correlations.

\begin{table*}
\centering                                                       
\caption{Profile parameter fit results}
\label{tab_sum}
\begin{tabular}{cccccc}                                      
\hline                                                           
\hline                                                           
Subgroup   & $\sigma_z$   & $\sigma_{\gamma}$ (X-ray) & $\sigma_{\gamma}$ (BGG) & $M_{200}$ (X-ray)          & $M_{200}$ (BGG)           \\
(X-ray ID) & (km s$^{-1}$) & (km s$^{-1}$)            & (km s$^{-1}$)          & ($10^{14} h^{-1} M_{\odot}$) & ($10^{14} h^{-1} M_{\odot}$) \\
\hline                                                           
2          & $303\pm 60$  & $240^{+90}_{-190}$        & $230^{+95}_{\dots}$       & $0.3^{+0.1}_{-0.2}$         & $0.2^{+0.2}_{-0.2}$            \\ 
3          & $406\pm 83$  & $530^{+45}_{-55}$         & $425^{+60}_{-70}$        & $1.1^{+0.4}_{-0.4}$         & $0.9^{+0.4}_{-0.3}$            \\ 
4          & $580\pm 100$ & $450^{+45}_{-50}$         & $445^{+45}_{-50}$        & $1.6^{+0.5}_{-0.4}$         & $1.6^{+0.5}_{-0.4}$            \\ 
5          & $567\pm 119$ & $355^{+55}_{-70}$         & $480^{+45}_{-50}$        & $0.8^{+0.4}_{-0.3}$         & $1.6^{+0.5}_{-0.4}$            \\ 
\hline                                                           
6          & \dots        & $325^{+85}_{-120}$         & $305^{+90}_{-135}$        & $0.7^{+0.6}_{-0.5}$         & $0.7^{+0.2}_{-0.1}$            \\ 
\hline
\end{tabular}
\end{table*}

The resulting $\sigma_{\gamma}$ values are given in Table \ref{tab_sum}. Consistent with the reconstructed mass map in Fig. \ref{ksinv}, we do not detect a very significant lensing signal around X-ray peak 2, barely exceeding the $68\%$ confidence limit.

The velocity dispersion associated with X-ray peak 1 is necessarily kept constant, as the peak lies outside the ACS mosaic. Upon inspection, it turns out that varying this parameter between $0 \le \sigma_1 \le 820$ km s$^{-1}$ does not alter the results by more than $10\%$ of the $68\%$ confidence interval for X-ray peak 2, which lies closest to peak 1. The effect is even smaller for the other groups.

Similar to our assessment of systematics for the mass map reconstruction, we repeated the fit to a control signal by changing the phase of the shear by $\frac{1}{2}\pi$. The results are consistent with a control signal of $g_{\mathrm{c}}\approx 0$. Because of its less favorable lensing geometry ($\left\langle\beta\right\rangle = 0.42$), the constraints for group 6 are weaker, although it is still detected at a significance of $\sigma\approx 1.6$.

Finally, we determined how much our results would be affected if the signal were boosted by a dilution factor of $1+(a+\sigma_a)/\theta =1+1.25/\theta$  for group member contamination, as discussed in Sect. 4.2, using a conservative $1\sigma $ upper limit. We find that this does not alter the results by more than $37\% $ of the $68\%$ confidence intervals, justifying our earlier approach.

We repeated the fit around the BGGs as tracers of the halo centers. The results are very similar, with the fitted values also given in Table \ref{tab_sum}. There is some difference with up to $2\sigma$ deviations between the results for peaks 3 and 5, where the separation between X-ray peak and BGG is also the largest. The quality of the fit, in terms of a reduced $\chi^2$ value, is the same.

\subsubsection{$M_{200}$}

In the same manner, we determined NFW profile parameters from the distortion pattern in the ACS field around the subgroups.

Weak lensing data of individual groups or low-mass clusters do not have sufficient signal-to-noise to provide useful constraints on $M_{200}$ and $c_{200}$ simultaneously. Therefore, we employed the mass-concentration relation given in \citet{Mmasscon}, restricting the fit to one free parameter, $M_{200}$. The results of these fits are summarized in Table \ref{tab_sum}, both for the X-ray centroids and BGGs as tracers of the halo centers.

\subsubsection{Scaling relations}

G05 showed that the subgroups were consistent with the local $T_X-\sigma_z$ relation \citep{XueWu2000}, a fact which did not change with more spectroscopic data in T09. Here we did not determine 1D velocity dispersions from the redshift distribution of group members, but assumed the projections of 3D halo models. Hence, we are not limited by group member identification. As mentioned before, group centroiding can be a problem.

\begin{figure}[h]                                                
\centering                                                       
\resizebox{\hsize}{!}{                                           
\includegraphics[angle=90]{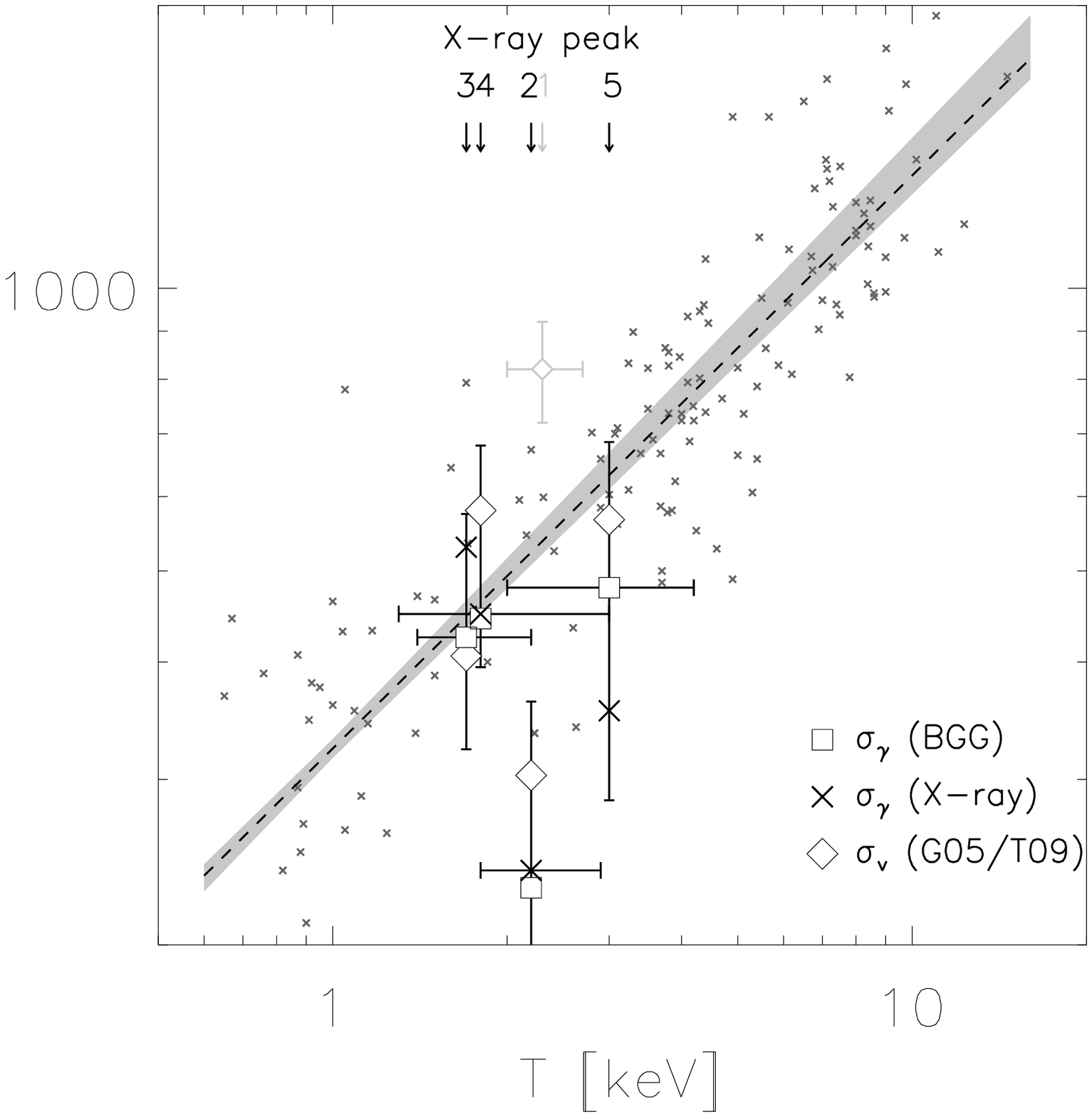}
\quad
\includegraphics[angle=90]{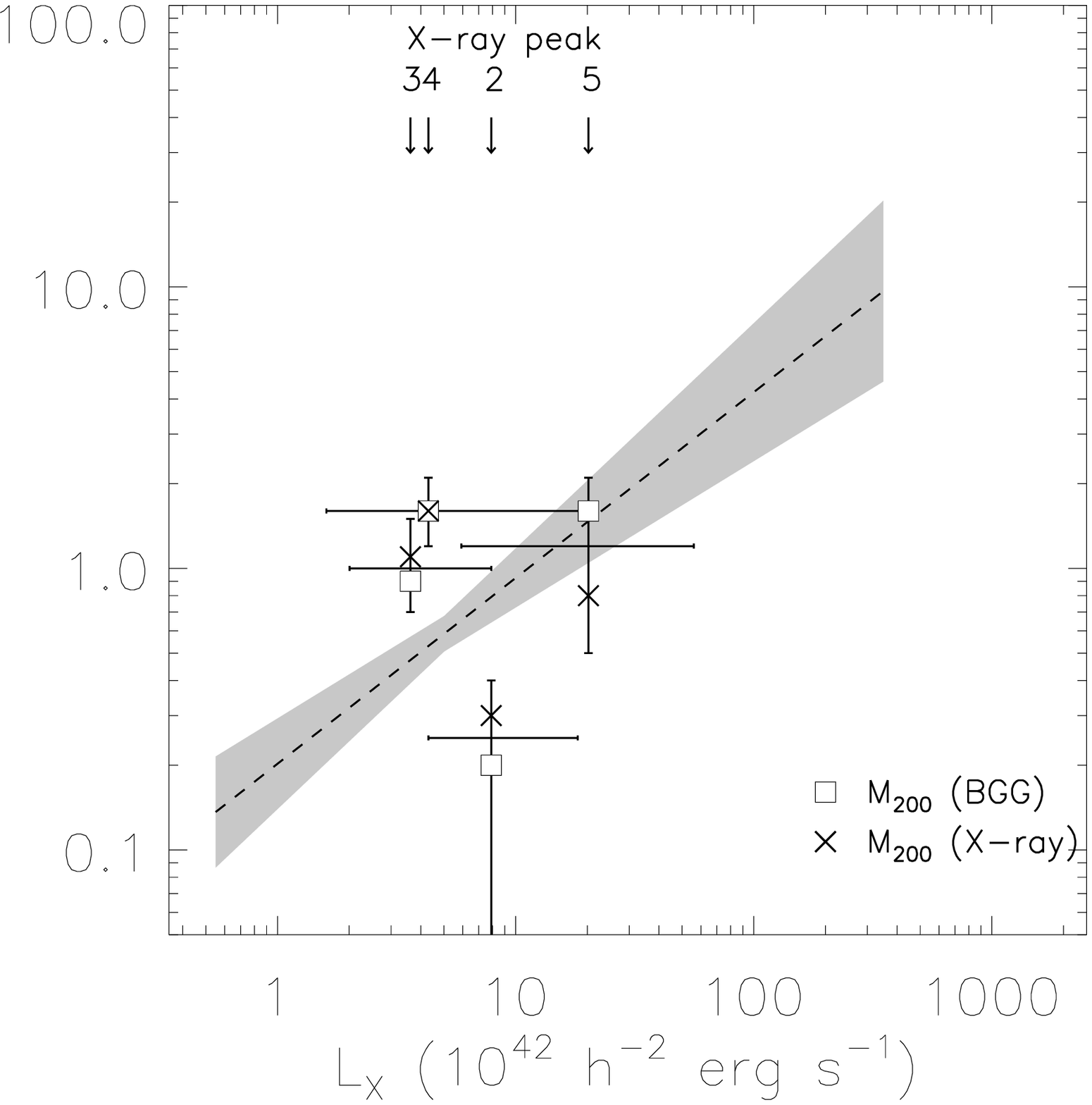}
}                                                                
\caption{Comparison of the properties of SG1120 with observed $\sigma - T$ \citep[][left]{XueWu2000} and $L_{\rm{X}} - M_{200}$ relations \citep[][right]{COSMOS}. Results based upon X-ray peaks and BGGs as center of mass are indicated by crosses and open squares, respectively, while velocity dispersions from G05 and T09 are shown as open diamonds. Horizontal error bars are plotted at the vertical median.}
\label{scale}                                                    
\end{figure}

Although the parameters of individual groups have shifted in this analysis, on average the groups still lie on the local $T_X-\sigma_z$ relation, showing a scatter of similar magnitude as the data in \citet[][Fig. \ref{scale}]{XueWu2000}.

\citet{COSMOS} constrained the $L_{\rm{X}} - M_{200}$ scaling relation using weak lensing data of groups in the COSMOS field. The supergroup as a whole is consistent with this scaling relation as well, within the scatter (Fig. \ref{scale}).

These interpretations would be reinforced if we did not take X-ray peak 2 into account and considered the conclusion in \citet{George12} that BGGs/MMGGs are better tracers of group halo centers than X-ray centroids.

Even though individual groups do not always lie precisely on the determined scaling relations, differences in environment and their effect on the astrophysical processes behind the observables used in these analyses create intrinsic scatter around these relations, which is averaged out in a stacking analysis such as employed in \citet{COSMOS}.

\section{Summary}

We have performed a weak lensing analysis of the coalescing supergroup SG1120 and showed that the underlying density distribution of matter is well traced by both visual galaxy light and X-ray emission. The subgroups of SG1120 have not yet interacted, but are expected to do so within short timescales, as projected separations are of about $1-4$ Mpc (G05). As such, the system is a unique demonstration of hierarchical structure formation.

Slight offsets between the peaks in the galaxy distributions, X-ray gas, and the total matter distributions are well within smoothing scales used and are consistent with an unrelaxed system on the verge of merging. We found that using either X-ray peaks or BGGs as tracers for the halo centers \citep{George12} has a minor impact on the derived halo parameters, with results consistent within $2\sigma$ error bars. We consider these conclusions to be an indication of the robustness of our results.

Furthermore, while the groups are close enough to be gravitationally bound (G05), the individual group halo masses are low enough compared to their separations to treat them as individual lenses, within parameter error bars.

The fitted profile parameters are consistent with well-demonstrated scaling relations, within the intrinsic scatter created by astrophysical variations \citep{COSMOS}. This is further confirmation that the observed structure of SG1120 is consistent with the paradigm of hierarchical structure formation, providing a unique example of this theoretical framework.

Structures such as SG1120 are rare. In fact, SG1120 should be seen as a single piece of a much larger puzzle, where confirmation from studies of similar structures is a necessity. The structure of SG1120 is uniquely oriented in the plane of the sky, and the subgroups show no signs of interaction yet, making it well suited to distinguish the various components and overdensities. An example of a well-studied heavier structure is the Cl 1604 supercluster \citep{CL1604_08}, where the complex structure presents difficulties in determining accurate masses, either using spectroscopic velocity dispersions \citep[e.g.,][]{CL1604_12} or weak lensing analyses of a few selected subclusters \citep{CL1604_05,CL1604_11}.

Especially the extension of studies like these to individual systems of lower mass like SG1120 will present a significant challenge, both in detecting such rare coalescing systems and in obtaining robust and accurate lensing measurements, given the lower S/N. An interesting approach is the combination of large existing spectroscopic group catalogs \citep[e.g.,][]{2PIGG,SDSS_Gr1,SDSS_Gr2,Robotham11} and recent or currently ongoing large sky imaging surveys of various width and depth, designed for lensing \citep[e.g.,][]{CFHTLenS,RCS2,KIDS} that are supported by extensive spectroscopic and color information.

\begin{acknowledgements}

We thank the anonymous referee for constructive and efficient comments that helped to improve this paper and the robustness of our conclusions. MS acknowledges support from the Netherlands Organization for Scientific Research (NWO). TS acknowledges support from the Netherlands Organization for Scientific Research (NWO), NSF through grant AST-0444059-001, and the Smithsonian Astrophysics Observatory through grant GO0-11147A. Observations taken by NASA HST G0-10499, JPL/Caltech SST GO-20683, and Chandra GO2-3183X3.

{\it Facilities.} VLT (VIMOS), VLT (FORS2), Magellan (LDSS3), HST (ACS), SST (MIPS), CXO (ACIS).

\end{acknowledgements}

\bibliographystyle{aa}
\bibliography{SG1120_lens.bib}

\appendix

\section{Quality flags and selection criteria for background sources.}

We assigned several quality flags to the source catalogs during detection and shape measurement.

We used the same rms noise model and deblending parameters as \citet{TS10} for object detection with {\tt SExtrator}. In addition to detection flags, we required at least eight adjacent pixels with values more than $1.4 \sigma$ above the background. We defined an initial S/N cut by flagging objects with {\tt FLUX\_AUTO}$/${\tt FLUXERR\_AUTO}$<10$.

We furthermore selected sources with a minimum size compared to the smearing induced by the PSF. We excluded sources for which the half-light radius $r_h$ \citep[as defined in][]{KSBp_Erben} compared to that of the average star is not smaller than $r_h > 1.2r_h^*$.

Finally, we selected sources with a KSB shape measurement S/N \citep[defined in][]{KSBp_Erben} larger than 4, to be consistent with KSB+ studies using a similar definition of the source S/N. In this pipeline, the effect of smearing and shearing by the PSF is for an important part described by the $P^g$ tensor. To avoid being dominated by noise, we excluded sources for which Tr$(P^g)/2 < 0.1$ \citep[see][for technical details and terminology]{KSBp_Erben}.

In the final source selection, the catalog of 8273 galaxies is reduced to 7012, $\sim 64$ galaxies/arcmin$^2$, that pass all quality criteria from detection and shape measurement.

\end{document}